# Tailoring mechanical properties and electrical conductivity of flexible niobium carbide nanocomposite thin films

Luis Yate,*[a] L. Emerson Coy,[b] Guocheng Wang,[a] Mikel Beltrán,[c] Enrique Díaz-Barriga,[d] Esmeralda M. Saucedo,[d] Mónica A. Ceniceros,[d] Karol Załęski,[b] Irantzu Llarena,[a] Marco Möller[a] and Ronald F. Ziolo[d]

Flexible NbC nanocomposite thin films with carbon content ranging from 0 to 99 at.% were deposited at room temperature on Si (100) and polystyrene substrates by non-reactive magnetron sputtering from pure Nb and C targets without applying bias voltage to the substrates. HRTEM images reveal that the films exhibit a nanocomposite structure consisting of NbC nanocrystals (2 to 15 nm in size) embedded in an amorphous carbon matrix. By simply adjusting the Nb flux in the plasma, we can monitor the nanocrystal size and the percent of free-carbon phase in the films, which in turn allows for the tailoring of both mechanical properties and electric conductivity of the films. It was found that the films composed of ~8-10% free-carbon exhibited a relative high hardness and elastic recovery, around 23 GPa and 85%, respectively, and an electrical conductivity of $2.2 \times 10^6$ S/m at 22ºC. This study indicates the potential of this non-reactive sputtering approach in depositing hard, elastic and electrically conductive nanocomposite films at low temperatures, which is especially useful for preparation of films on temperature sensitive polymers or plastic substrates for nano- and micro- electronics applications.

## 1. Introduction

In the recent years there has been an increased demand for stable, low-cost, electrically conductive materials for flexible electronics applications. Many oxide [1,2] and graphene [3,4] materials have been studied for this purpose because of their well-known electronic properties and potential suitability. In many applications, however, the mechanical properties, such as hardness, elastic modulus and elasticity, are often overlooked as is the interactive chemical environment to which the materials are subjected. Properties, such as tensile or compressive stresses during the flexions, and tribology are especially important if the surface of the materials is in contact with other surfaces. $CN_x$ films in this regard are recognized to have a very high elasticity or elastic recovery of almost 100% [5,6], but have relatively low hardness from 7 to 15 GPa [6] and conductivities in the order of $10^{-9}$ to $10^5$ S/m depending on the nitrogen concentration [7].

Transition metal carbides (MeC or Me-DLC), on the other hand, constitute a scientifically and technologically formidable class of materials which are receiving increased attention because of their mechanical, chemical, electronic and optical properties, with potential applications in nanoscale systems [8–11]. Recent activity in the MeC systems has focused on the synthesis of 2D materials, called MXene structures in analogy to graphene [12], and as nanocomposites, nanoparticles, tubes and rods [13–17].

In particular, nanocomposite hard coatings of transition metal carbides, have been widely investigated due to the aforementioned properties, their superior toughness, biocompatibility, excellent tribological properties and lower residual stresses than that of conventional pure diamond-like carbon (DLC) films [18–21]. Nanocomposite coatings have improved hardness and elasticity because of their unique structure with nanocrystals embedded in an amorphous matrix, where the fine control of the amorphous phase content is critical. The amorphous matrix can accommodate higher internal stresses and inhibit dislocations from passing through the amorphous grain boundaries [22,23].

Many studies have been reported on the dependence of the nanocomposite structure on the growth conditions. Most of the studies have been dedicated to the formation of TiC, CrC, WC and SiC nanocrystals in amorphous or hydrogenated amorphous carbon matrix, a-C and a-C:H, respectively [21,24–29]. Little work has appeared, however, on niobium carbide, NbC, which has an outstandingly high melting temperature among the carbides of ~3610ºC [14], higher electrical conductivity than that for VC, HfC, ZrC or TiC and superconductivity at around 10 K [30]. Additionally, NbC has also shown better catalytic activity than that of other carbides [31]. These, and the aforementioned properties, have inspired new research and interest in the niobium carbides and open the possibility to extend the already wide range of potential applications of the niobium carbides in new areas of nanoscience and nanotechnology.

The benefits of non-reactive deposition in magnetron sputtering, i.e., enhanced ion bombardment and the inhibition of the formation of hydroxide species, were recently reported [32] and were utilized in this study. Here, we present a thorough study of the deposition of NbC films with carbon contents ranging from 0 to ~100% deposited using non-reactive magnetron sputtering and, importantly, without applying any substrate temperature or bias voltage to substrates. Under these conditions, we have determined the effect of varying the Nb flux in the plasma on the chemical composition and physical properties of the films by changing the applied r.f. power to the Nb target and demonstrate the effective deposition of such films on flexible polymer substrates at room temperature.

**Table 1**. Relationship of the r.f. power applied to the Nb target and deposition rate as well as chemical composition of the films by XPS analyses and nanocrystal size by HRTEM.

| Sample | Nb r.f target power (W) | Deposition rate (μm/h) | Nb (at. %) | C (at. %) | O (at. %) | Free C-C content (%) | Nanocrystal size (nm) |
|---|---|---|---|---|---|---|---|
| Pure C | 0 | 0.08 | 0.0 | 99 | 1 | 99 | - |
| NbC20 | 20 | 0.08 | 9 | 86 | 5 | 82 | ~2 |
| NbC50 | 50 | 0.13 | 37 | 56 | 7 | 30 | 2-5 |
| NbC100 | 100 | 0.13 | 52 | 39 | 9 | 9 | 5-8 |
| NbC230 | 230 | 0.32 | 59 | 30 | 11 | 5 | 10-15 |
| Pure Nb | 230 | 0.28 | 72 | 10 | 18 | 0 | >10 |

## 2. Experimental

Flexible NbC films were deposited on silicon (100) and flexible polystyrene (Thermanox® coverslips) substrates with non reactive magnetron sputtering using an AJA-ATC 1800 system with a base pressure of $10^{-7}$ Pa. The deposition of the films was done with two separate 2 inch elemental targets, with a purity of 99.999% for carbon (graphite) and 99.95% for Nb, in a confocal configuration at a pressure of 0.25 Pa of pure Ar. The carbon and niobium targets were acquired from Demaco-Holland and AJA international-USA, respectively. The samples were grown under floating potential at room temperature by turning off both the bias power source and the substrate holder heating facility during depositions. The distance between target and substrates was about 15 cm. Prior to deposition, the substrates were sputter cleaned with a negative bias of 180 V (25 W) in a 4 Pa Ar atmosphere for 10 min. In order to improve the adhesion of the films to the substrates, a pure Nb layer of ~30 nm was deposited onto the substrates at a Nb target r.f. power of 230 W. NbC films with different compositions and thicknesses between 0.2 and 0.3 μm were obtained by varying the Nb target r.f. power (0, 20, 50, 100 and 230 W and hereafter referred to as Pure C, NbC20, NbC50, NbC100 and NbC230, respectively) while keeping the carbon target d.c. power constant at 380 W. A pure niobium film was also deposited by applying a r.f. power of 230 W to the Nb target, Table 1.

Film thicknesses and cross-section morphology were determined by SEM with a JEOL JSM-6490LV and the deposition rate was obtained dividing the thickness by the deposition time. X-ray photoelectron spectroscopy (XPS) analyses were performed by means of a SAGE HR100 (SPECS) with a non-monochromatic source (Mg Kα 1283.6 eV) after a short etching of the sample surface with Ar$^+$ ions at 3 kV energy in order to remove contamination. According to the studies of Lewing et al. [33] the sputter damage contribution caused by the Ar+ ion energy of around 3 kV in the analogous TiC nanocomposite system does not exceed more than 10% of the total signal. The chemical composition of the samples was obtained from the areas of the peaks using the software CasaXPS V2.3.15dev87. Raman spectra of the films were recorded with use of a Renishaw in Via Raman spectrometer in the 500 cm$^{-1}$ to 2500 cm$^{-1}$ range using a green laser light source at a wavelength of 532 nm.

Grazing incidence X-ray diffraction (GI-XRD) analysis was conducted for structural investigations of the coatings using an X'Pert PRO PANalytical X-ray diffractometer with Cu Kα radiation at 0.5° incidence angle. Samples for cross-section HRTEM were prepared by a dual beam, focused ion beam (FIB) system, Workstation Quanta 200 3D. The ion source used for the FIB preparation was gallium at 30 kV of energy for the rough cuts and 10 kV of energy for the fine cuts. High resolution cross-section images were obtained using a FEI Titan 80-300 electron microscope working at 300 kV.

Nanohardness, elastic modulus and elastic recovery of the coatings were measured by nanoindentation (Hysitron TI 950 TriboIndenter) using a Berkovich diamond indenter at different loads. Hardness and elastic modulus values were determined from the load-displacement curves by the Oliver–Pharr method [34]. The elastic recovery of the samples was determined as the percentage of the residual imprint compared to the total displacement of the load-displacement curves. Cross-sections of the nanoindentations were prepared with a JEOL-4000 FIB. The ion source used was gallium at 30 kV of energy. The TEM images of the cross-sectioned nanoindentations were acquired with a JEOL-TEM 1400 electron microscope working at 120 kV.

Average electrical conductivity values were calculated according to the van der Pauw method[35] from 3 sheet resistivity measurements made on each sample at 22ºC using a Keithley 2400 SourceMeter with I = 100 mA and a 4-point probe configuration with gold wires attached to the films with silver paint. The film thicknesses used to calculate the average conductivity values were the average thickness values determined from both FIB sectioned films and cross-section SEM images.

## 3. Results and Discussion

A cross-section SEM image of the NbC230 sample is shown in Fig. 1a. The image shows the film with a very dense and featureless structure with no discontinuities or voids between the film and the silicon substrate.

The thickness values were determined from the cross section SEM images of all samples and the deposition rates, shown in Table 1, were calculated dividing the thickness by the deposition time. The deposition rate of the films was strongly dependent on the r.f. power applied to the Nb target and was found to vary from 0.08 to 0.3 μm/h as the Nb flux was increased.

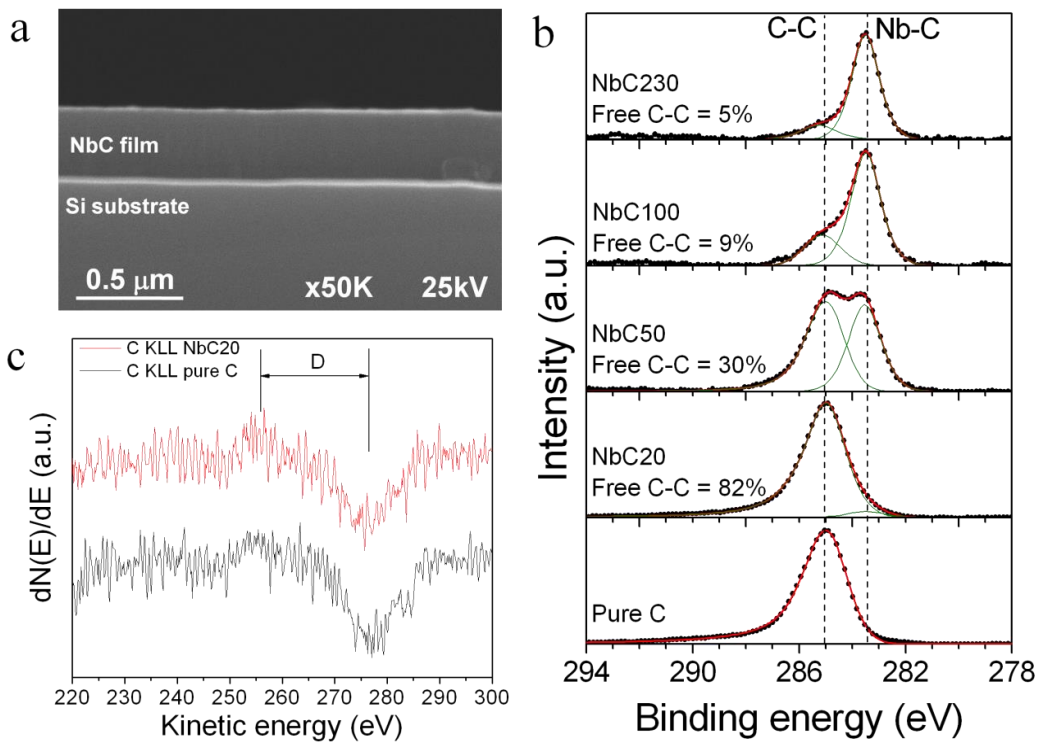

**Fig. 1. a)** Cross-section SEM image of the NbC230 sample, showing a very dense, featureless and well adhered structure, **b)** Fitted XPS C 1s spectra of the NbC and pure C films, **c)** First derivative C KLL spectra of the Pure C and NbC20 samples.

The analysis of chemical composition of the samples and the identification of the different chemical bonding states of the carbon atoms were done using XPS. The compositions of the films determined, based on the relative areas, *A*, and sensitivity factors, *S*, of the C 1s, Nb 3d and O 1s peaks using equation (1)[36], are shown in Table 1.

$$C_i = \sum(A_i/S_i)/(A_j/S_j) \quad (1)$$

As expected, the Nb content in the films increased with the r.f. power applied to the Nb target. The range of carbon content found in the films varied from almost 0% for the pure Nb film to almost 100% for the pure C film. The pure Nb film showed around 10% of adventitious carbon on the surface. Samples with the higher niobium content also showed the higher content of oxygen. This can be attributed to the very high affinity of metals for oxygen. In non-stoichiometric metal carbides, $MeC_{1-x}$, for example, empty octahedral interstitial sites are available for stray oxygen atoms [17].

The high resolution C 1s carbon spectra of the NbC and Pure C films are shown in Fig. 1b. The C 1s spectra of the NbC films show the presence of two different carbon chemical states, one associated to C-C bonds at around 285.0 eV [37] and another one associated to Nb-C bonds at around 283.2 eV [38,39], suggesting the presence of amorphous carbon and niobium carbide phases. The occurrence of Nb-C bonds indicates the chemical reaction at room temperature of Nb and C on the surface of the substrates. The C 1s spectra in Fig. 1b shows that there is a clear evolution in the carbon bonding environment with the increasing Nb content in the films as evidenced by the increase of Nb-C bonds. The spectrum of the Pure C film was fitted with an asymmetric shape characteristic of a pure graphite-like carbon structure. The spectra of the NbC films were fitted with an asymmetric shape for the C-C component and a symmetric Gauss-Lorentz shape for the Nb-C component, except for the NbC100 and NbC230 samples, which were fitted with two symmetric peaks.

The content of the free carbon content, free C-C, in the different films was calculated from the fitting of the C 1s spectra and the results are shown in Table 1. The free C-C content varied over a wide range, from 5% in the NbC230 sample to 99% in the Pure C sample.

The component associated with the C-C bonds in the C 1s spectra contains information of the $sp^2$ and $sp^3$ content, the peaks associated to each bonding configuration, however, are too close and are difficult to separate by fitting [29]. Moreover, it has been previously reported that the C 1s peak of samples with $sp^2$ contents from 0 (diamond) to 75% (pyrolytic graphite) did not exhibit significant variation in shape as the $sp^2$ content varied [40]. In order to identify the effect of the addition of the niobium on the $sp^2/sp^3$ carbon content, a deeper XPS analysis was conducted by measuring the C KLL spectra as proposed by Turgeon et al. [40] and Mezzi et al. [41]. This method consists of measuring the separation in energy between the most positive maximum and most negative minimum in the first derivative C KLL spectra, called the D-parameter, and then the $sp^2$ content is related to an approximate linear dependency with the D-parameter and the values are compared to those of Turgeon et al. [40] and Mezzi et al. [41].

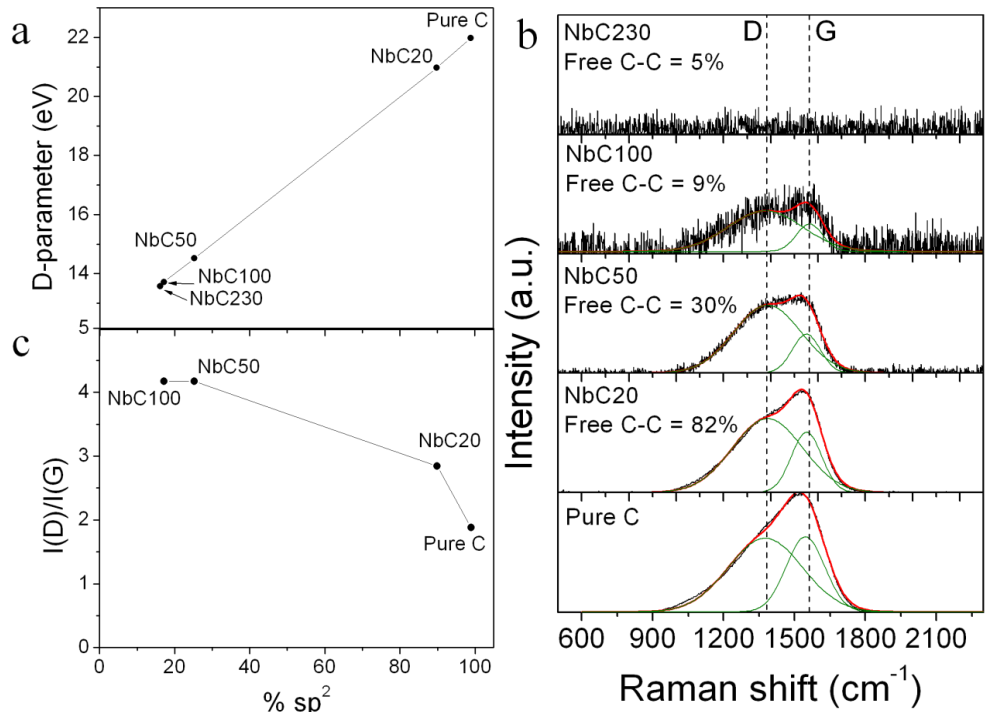

**Fig. 2**. **a)** D-parameter vs. the $sp^2$ content for Pure C and NbC samples, **b)** Fitted Raman spectra of the Pure C and NbC samples, **c)** I(D)/I(G) ratio obtained from Raman analysis vs. the $sp^2$ content obtained from the XPS analysis.

However, it should be stated that these values are only semi-quantitative and gives relative values, since a calibration curve has to be done for each different set of materials. The first derivative of the carbon C KLL spectra for the Pure C and NbC20 samples are shown in Fig. 1c.

The D-parameter against the $sp^2$ content is plotted in Fig. 2a. The Pure C sample presents a graphite-like structure as demonstrated by its elevated $sp^2$ content. As the Nb content in the films is increased with the r.f. power from sample NbC20 to NbC230, the $sp^2$ content decreases, indicating a transformation in the carbon bonding environment from a graphite-like to an amorphous-like structure.

The free amorphous carbon structure was further studied by Raman spectroscopy. Raman spectra of the Pure C and the different NbC samples are shown in Fig. 2b. Spectra of the films with free C-C content higher than 9% shows the characteristics D peak at around 1380 cm$^{-1}$, which is attributed to disordered $sp^2$ carbon and the G peak at around 1580 cm$^{-1}$, attributed to graphitic carbon. The presence of the D and G peaks suggests the existence of an amorphous carbon matrix [42]. However, as the free C-C content decreases to 5% in the NbC230 sample, no Raman signal corresponding to the free amorphous carbon matrix can be detected.

Indirect information of the $sp^2/sp^3$ hybridization content may be obtained from the I(D)/I(G) ratio, defined as the ratio of the areas of the D and G peaks, as obtained from the fitting of the Raman spectra [42,43]. Fig. 2c presents the calculated I(D)/I(G) ratio correlated to the $sp^2$ content obtained from the analysis of the first derivative C KLL spectra. The decreasing of the I(D)/I(G) ratio with the $sp^2$ content indicates an amorphization of the carbon according to previous studies in amorphous carbon reported by Ferrari et al. [42].

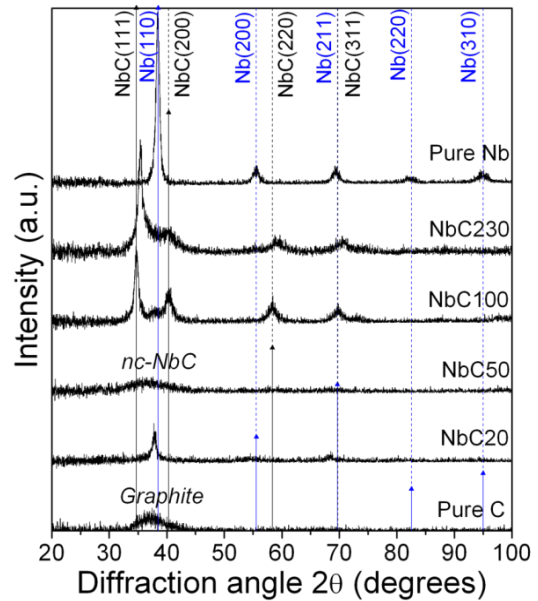

**Fig. 3**. GI-XRD patterns of the Pure C, Pure Nb and NbC films.

Fig. 3 shows the GI-XRD patterns of the Pure C, Pure Nb and NbC films. The pattern for the Pure C film shows only a broad band centred at around 2θ = 38° characteristic of a graphite like/amorphous carbon structure. The XRD pattern of the NbC20 sample shows peaks associated to the cubic BCC phase of Nb that can be associated to the Nb adhesion layer.

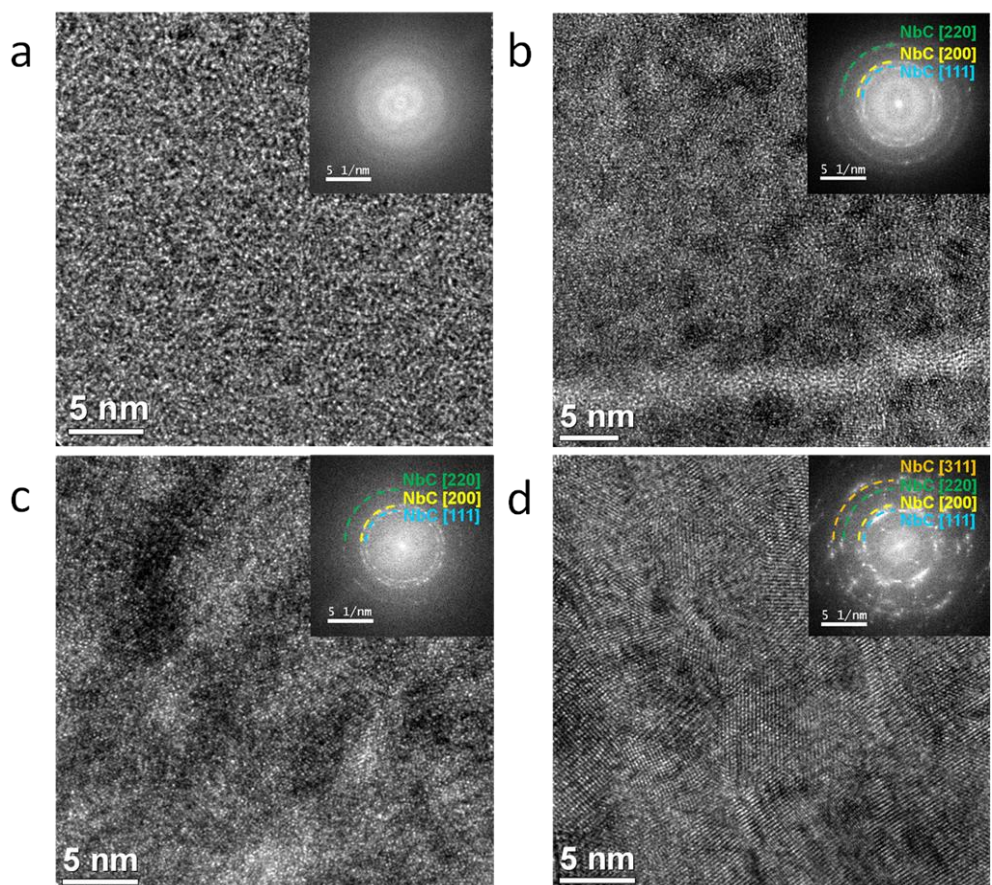

**Fig. 4**. Cross-section HRTEM images for the different NbC films **a)** NbC20, **b)** NbC50, **c)** NbC100 and **d)** NbC230. Fast Fourier transform patterns are shown as insets.

The NbC50 sample, deposited at a higher r.f. power applied to the Nb target, presents a broad and asymmetric band centred around 2θ = 36º that can be attributed to the formation of nanocrystalline/amorphous niobium carbide. The NbC100 sample, presents the crystalline structure of cubic Fm-3m NbC (JCPDS-ICDD 38-1367). In this sample, the Nb ions have enough energy to bond to carbon in the plasma and to form a crystalline structure. The NbC230 sample shows a higher crystallinity probably related to the even higher energy of the Nb ions and also to the lower content of the free C-C in the matrix. Finally, the pattern for the Pure Nb film shows a very well defined crystalline structure attributed to cubic BCC Nb (JCPDS-ICDD 34-0370).

The XRD results for the NbC films demonstrate that a higher flux of niobium in the plasma, generated by a higher r.f. power applied to the Nb target, results in a higher average kinetic energy of the ions impinging on the growing film favouring the formation of Nb-C bonds, as demonstrated with XPS analysis, and the NbC crystalline structure.

The occurrence of the cubic NbC structure with a relatively good crystallinity and without mixing with other phases in the films deposited at room temperature and without applying bias voltage contrast to previously reported studies where $Nb_2C$ and NbC mixed phases were obtained [39,44–48].

Fig. 4 shows the cross-section HRTEM images of the different NbC films. The HRTEM images demonstrate the nanocomposite structure of the films, where NbC nanocrystals of different sizes were embedded in an amorphous carbon matrix. It is clearly seen that the microstructure of the films is strongly dependent on the Nb content, which can be tuned by varying the r.f. power applied to the Nb target. The NbC20 sample (Fig. 4a) with a content of free C-C of 82%, determined by XPS analysis, shows very small nanocrystals, probably made of Nb or NbC, with sizes around 2 nm embedded in the carbon matrix. The Fast Fourier transform (FFT), inset of Fig. 4a, shows some diffuse rings confirming the nanocrystalline/amorphous structure of the sample. In the NbC50 sample, the free C-C content is lower, 30%, and the HRTEM image shows the presence of NbC nanocrystals with sizes of about 4 nm (Fig. 4b). The FFT pattern inset in Fig. 4b, shows sharper rings attributed to the [111], [200] and [220] of the cubic NbC structure, which is consistent with the XRD results. The bright line in Fig. 4b was caused by the Nb plasma briefly off for less than 60 seconds during deposition. As the Nb r.f. power target was increased to 100 W, the size of the NbC nanocrystals grew to 5-6 nm, as shown in Fig. 4c, and the diffraction rings assigned to the cubic NbC structure became more clearly defined (inset of Fig. 4c). Distinguished from the other films, the NbC230 film (Fig. 4d) shows a microstructure with larger and elongated NbC nanocrystals of around 10 nm in size. Besides the [111], [200] and [220] diffraction rings, an additional ring assigned to [311] planes of the cubic NbC structure also appears in the FFT pattern of the NbC230 film, as shown in the inset of Fig. 4d, indicating an even higher crystallinity of the film, compared to that of the other films, and is in agreement with the XRD results

(Fig. 3).

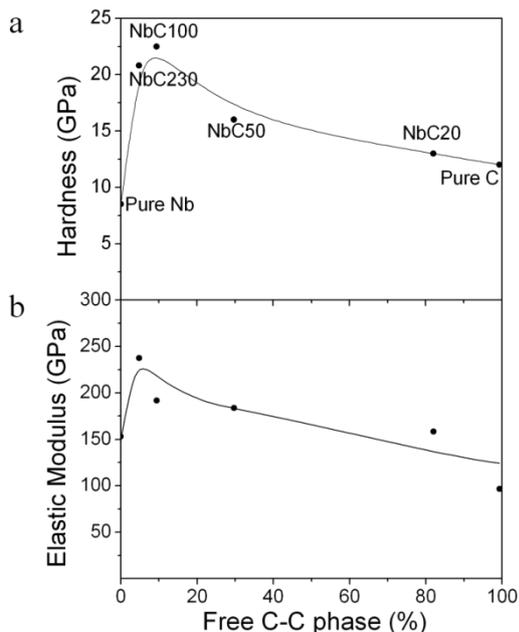

**Fig. 5.** **a)** Nanohardness and **b)** Elastic modulus of the NbC films as a function of the free C-C phase. The solid lines are meant as a guide to the eye.

Figs. 5a and 5b show the nanohardness and elastic modulus of the NbC films as a function of the free C-C content, respectively. As the free C-C content in the films increases, both the nanohardness and the elastic modulus increase at first, until reaching a maximum for the NbC100 film with a free C-C content of 9%, then gradually decrease, which is the classical behaviour of a nanocomposite material described by Veprek et al. [22,23]. Subsequent increments of the free C-C phase produced a decrease in the size of the NbC nanocrystals, as seen in the HRTEM analysis, and a decrease in the nanohardness and elastic modulus values until reaching the value for pure carbon.

The nanohardness of the films, from 10 to 23 GPa, is similar to that reported by (a) Nedfors et al. for nanocomposite NbC films deposited by non-reactive magnetron sputtering [49], (b) Zhang et al. for NbC-H films deposited by reactive magnetron sputtering of a Nb target in $CH_4$ atmosphere [48], (c) Benndorf et al. for NbC-H films deposited by reactive magnetron sputtering in $C_2H_2$ atmosphere [38] and (d) Zou et al. for NbC films obtained by a chemical solution technique [50]. However, none of these reports mentioned exploring the full range of compositions for the NbC nanocomposite systems, from 0 to almost 100% of the free C-C content.

As defined by Leyland et al. [51], the ratio between the nanohardness and the elastic modulus (H/E) can be used as an indication of the wear resistance of a nanocomposite coating. In this study the H/E ratio of the NbC films varied from 0.05 for the Pure Nb film to 0.12 for the NbC100 sample and the Pure C sample composed of mostly 100% of pure $sp^2$ graphitic carbon. This suggests that the NbC100 sample with the higher nanohardness could present a tribological behaviour similar to that of diamond-like carbon (DLC) films, which have been very well studied and recognized for their good wear properties [52].

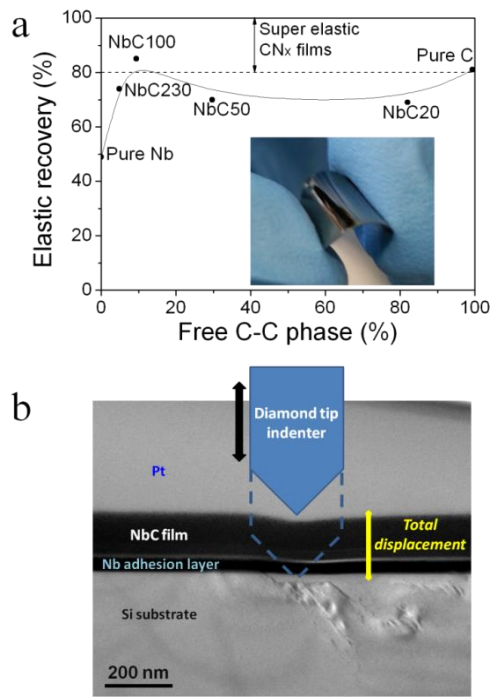

**Fig. 6.** **a)** Percent elastic recovery of the NbC films as a function of the free C-C phase. The solid line is meant as a guide to the eye. The region of super elastic $CN_x$ films is shown for comparison; The inset shows one of the NbC films deposited on a polystyrene flexible substrate being flexed, and **b)** Cross-section TEM image of the NbC100 sample after an indentation experiment. The image confirmed the high elastic recovery of the film after the indenter penetrated about 200 nm into the sample while the silicon substrate showed multiple fractures and dislocations. The dashed line in the image indicates the total penetration of the indenter into the sample. The Pt film was added post indentation for cross-sectioning.

We note with interest that the % elastic recovery for the NbC100 sample, Fig. 6a, falls within the super elastic regime similar of that reported for fullerene like $CN_x$ films, with elastic recoveries between 80 and 100% [5,6]. For the Pure C film, with a graphite-like structure, the elasticity depends strongly on the degree of order. However, as in the case of the fullerene like $CN_x$ films, the high elastic recovery for the Pure C film, 81%, is generally accompanied with a low nanohardness of ~12 GPa. On the other hand, the nanocomposite structure of the NbC100 sample gives both high elastic recovery, 85%, and a high hardness, 23 GPa. The inset in Fig. 6a, shows a representative NbC100 film deposited on a polystyrene substrate being flexed. Preliminary studies by cross-section TEM of one indentation performed in the NbC100 sample confirmed the high elastic recovery of the film after the indenter penetrated about 200 nm into the sample while the silicon substrate below the film showed multiple fractures and dislocations, Fig. 6b. As in the case of other nanocomposite structures, the super elastic behaviour in the NbC100 sample is attributed to the NbC nanograins restricting excessive deformation in the amorphous carbon matrix [53].

The samples were well adhered to the substrates and did not delaminate after the scotch tape test which was also used to evaluate the adhesion strength of other carbon-based films [54]. The high elasticity and adhesion of the present NbC films deposited on various polymer substrates enables the formation of room-temperature deposited, electrically conducting NbC elastic films with potential applications in such areas as flexible electronics, biomedical stents, and mechanical and chemical wear-resistant applications.

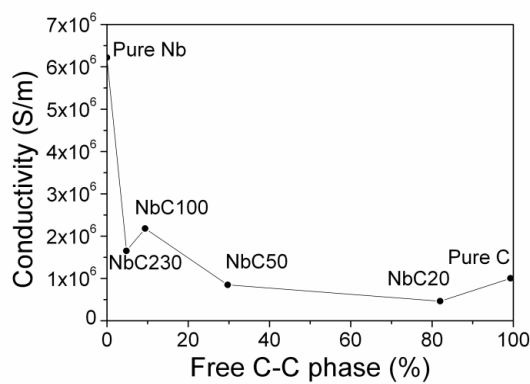

**Fig. 7**. Electrical conductivity of the NbC films as a function of the free C-C phase. The solid line is meant as a guide to the eye.

The electrical conductivities of the NbC films as determined from 4 point probe measurements of the sheet resistance at 22ºC are plotted in Fig. 7 as a function of the free C-C phase. NbC100 is the most conductive of the NbC films at room temperature and shows a promising potential for flexible electronics applications. Although the potential value for applications in industrial and medical devices for these types of hard, flexible films deposited at room temperature is evident, we note that the coupling of such films with flexible photovoltaic films offers potential value for light weight deployable solar cells used to power satellites and to form films that are being developed for textile and outerwear use.

## Conclusions

Flexible NbC nanocomposite thin films were deposited by non-reactive magnetron sputtering from pure Nb and C targets without applying any substrate temperature or bias voltage to the silicon and flexible polystyrene substrates. The carbon content in the samples varied from 0 to almost 100 at.%. It was found that the carbon and niobium content and the carbon bonding environment can be controlled through the r.f. power applied to the Nb target. The increase in Nb content led to an increase in the proportion of Nb-C bonds and to the change from a graphite-like to an amorphous-like structure in the carbon matrix. The crystallinity and the size of the NbC nanocrystals in the nanocomposite can be tailored by the r.f. power applied to the Nb target. The nanohardness of the sample was strongly correlated to the carbon content and structure of the samples, showing the typical behaviour of a nanocomposite structure.

The non-reactive deposition approach without the application of external energy, like temperature or bias voltage, represents a green approach to the preparation of NbC nanocomposite structures with outstanding properties. The room temperature deposition of hard, elastic and electrically conductive nanocomposite NbC films on organic substrates broadens the possibility for new and cost-effective applications in the active fields of energy conversion and flexible electronics.

## Acknowledgements

L.E. Coy thanks the financial support from the National Centre for Research and Development under research grant "Nanomaterials and their application to biomedicine", contract number PBS1/A9/13/2012. We thank Layza A. Arizmendi and Gilberto Hurtado at CIQA for assistance with TEM imaging and for electrical measurements.

## Notes and references

[a] CIC biomaGUNE, Paseo Miramón 182, 20009 Donostia-San Sebastian, Spain; Tel: +34 943005337; E-mail:lyate@cicbiomagune.es
[b] NanoBioMedical Center, Adam Mickiewicz University, Umultowska 85, 61-614 Poznan, Poland; E-mail: coyeme@amu.edu.pl
[c] CIC nanoGUNE, Tolosa Hiribidea 76, 20018 Donostia-San Sebastian, Spain
[d] Centro de Investigación en Química Aplicada (CIQA), 25294 Saltillo, Coahuila, México